\documentclass[a4paper,10pt]{article}
\usepackage{epsfig}
\usepackage{amssymb}
\begin{document}
\thispagestyle{empty}

\newcommand{\etal}  {{\it{et al.}}}  
\def\Journal#1#2#3#4{{#1} {\bf #2}, #3 (#4)}
\def\PRD{Phys.\ Rev.\ D}
\def\NIMA{Nucl.\ Instrum.\ Methods A}
\def\PRL{Phys.\ Rev.\ Lett.\ }
\def\PLB{Phys.\ Lett.\ B}
\def\EPJ{Eur.\ Phys.\ J}
\def\IEEETNS{IEEE Trans.\ Nucl.\ Sci.\ }
\def\CPCD{Comput.\ Phys.\ Commun.\ }

\smallskip

\bigskip
\bigskip

{\huge\bf
\begin{center}
Bose-Einstein correlations and the  stochastic scale of light hadrons emitter source
\end{center}
}

\vspace*{\fill}

\begin{center}
{\LARGE\bf  G.A. Kozlov}

\vspace*{\fill}
\noindent
 {\large
  Bogolyubov Laboratory of Theoretical Physics\\
  Joint Institute for Nuclear Research \\
   Joliot Curie st.6, Dubna, 141980  Russia\\
 }
\end{center}
\vspace*{\fill}

 \section*{Abstract}
\noindent
Based on quantum field theory at finite temperature we carried out
new results for two-particle Bose-Einstein correlation (BEC) function ${C_2}(Q)$
in case of light hadrons.
The important parameters of BEC function related to the size of the emitting source,
mean multiplicity, stochastic forces range with the
particle energy and mass dependence, and the temperature of the source
are obtained  for the first time.
Not only the correlation between identical
hadrons are explored but even the off-correlation between non-identical particles
are proposed. The correlations of two bosons in 4-momentum space presented
in this paper offer useful
and instructive complimentary viewpoints to theoretical and experimental works in
multiparticle femtoscopy and interferometry measurements at hadron colliders.
This paper is the first one to the next opening series of
works concerning the searching of BEC with experimental data where the parameters
above mentioned will be retrieved.

\vspace*{\fill}


\section{Introduction}

\bigskip

For the aim to explore the correlations of Bose-Einstein type (BEC)
one needs to use the properties of a particle detector, e.g., its
tracking system to study the hadron processes at some energy region.
Such a study will be done soon in the next papers.

This paper describes an attempt to address the problems of BEC
within the theoretical aspects prior the real data will be analyzed.

Over the past few decades, a considerable number of studies have
been done on the phenomena of multi-particle correlations observed
in high energy particle collisions (see the review in [1]). It is well
understood that the studies of correlations between produced
particles, the effects of coherence and chaoticity, an estimation of
particle emitting source size and the temperature play an important
role in this branch of high energy physics.

By studying the Bose-Einstein correlations of identical
particles (e.g., like-sign charge particles of the same sort)
or even off-correlations with respect to different-charge bosons,
it is possible to predict and even  experimentally
determine the time and spatial region over which particles do not
have the interactions. Such a surface is called as decoupling one.
In fact, for an evolving system such as, e.g.,  $p \bar p$ collisions, it is
not really a surface, since at each time there is a spread out
surface due to fluctuations in the final interactions, and the shape
of this surface evolve even in time. The particle source is not
approximately constant because of energy-momentum conservation
constraint.

More than half a century ago Hanbury-Brown and Twiss [2]  used
BEC between photons to measure the size of distant stars. In the
papers in [3] and [4] , the master equations for evolution of
thermodynamic system  created at the final state of  the
(very) high multiplicity process were established. The equations
have the form of the field operator evolution equation
(Langevin-like [5]) that allows one to gain  the basic
features of the emitting source space-time structure. In particular,
it has been conjectured and further confirmed that the BEC is strongly
affected by non-classical off-shell effect.

The shapes of BEC function were experimentally established in the
LEP experiments ALEPH [6], DELPHI [7] and OPAL [8], and ZEUS Collaboration
at HERA [9], which also indicated a dependence of the measured so-called
correlation radius on the hadron $(\pi,\ K)$ mass. The results for
$\pi^{\pm}\pi^{\pm}$ and $\pi^{\pm}\pi^{\mp}$ correlations with $p\bar p $
collisions at $\sqrt s = 1.8$ TeV were published by E735 Collaboration in [10].

One of the aims of this paper is to carry out the extended model of BEC
in the framework of quantum
field theory at finite temperature $({QFT}_\beta)$ approach which to be
applied later to real experimental data on two-particle BEC.
It is known that the effective temperature of the vacuum or the ground state
or even the thermalized state of particles distorted by external forces is
occurring in models quantized in external fields. One of the main parameters
of the model considered here is the temperature of the particle source under the random
source operator influence.

Among the results obtained in this paper we mention a theoretical estimate accessible
to experimental measurements of two-particle BEC and proof that
quantum-statistical evolution of particle-antiparticle correlations are not
an artifact of the standard formalism but a quite general properties of
particle physics. The effect (called as surprized one) for non-identical particles
correlations was predicted already in [11].

\section{Two-particle BEC}
\label{bec}

A pair of bosons with the mass $m$ produced incoherently (in ideal
nondisturbed, noninteracting cases) from an extended source will
have an enhanced probability $C_{2}(p_{1},p_{2})=
N_{12}(p_{1},p_{2})/[N_{1}(p_{1})\cdot N_{2}(p_{2})]$ to be measured
(in terms of differential cross section $\sigma$), where
\begin{equation}
\label{e31}
N_{12}(p_{1},p_{2})=\frac{1}{\sigma}\frac{d^{2}\sigma}{d\Omega_{1}\,d\Omega_{2}}
\end{equation}
to be found close in 4-momentum space $\Re_{4}$ when detected
simultaneously, as compared to if they are detected separately with
\begin{equation}
\label{e32}
 N_{i}(p_{i})=\frac{1}{\sigma}\frac{d\sigma}{d\Omega_{i}}, \,\,\,
 d\Omega_{i}=\frac{d^{3}\vec p_{i}}{(2\pi)^{3}\,2E_{p_{i}}}, \,\,
 E_{p_{i}}=\sqrt {\vec p_{i}^{2}+m^{2}},\,\,\,
 i = 1, 2.
\end{equation}

The following relation can be used to retrieve the BEC function
$C_2(Q)$:
\begin{equation}
\label{Konstrukcia C2}
 C^{ij}_2(Q) = \frac{N^{ij}(Q)}{N^{ref}(Q)}, \,\,\,\,\,\,
 i,\, j= +,\, -,\, 0,
\end{equation}
where $N^{ij}(Q)$ in general case refer to the numbers
$N^{\pm\pm}(Q)$ for like-sign charge particles (eg.,
$\pi^{\pm}\pi^{\pm},\  K^{\pm} K^{\pm},\ \ldots$); $N^{\pm\mp}(Q)$
--- for different charge bosons (eg., $\pi^{\pm}\pi^{\mp},\  K^{\pm}
K^{\mp},\ \ldots$) or even for neutral charge particles $N^{00}(Q)$
(eg., $\pi^{0}\pi^{0},\  K^{0} K^{0},\ \ldots$) with
\begin{equation}
\label{eq_01}
 Q = \sqrt {-(p_1-p_2)_{\mu}\cdot (p_1-p_2)^{\mu}}= \sqrt{M^{2} - 4\,m^{2}}.
\end{equation}
In formula  (\ref{Konstrukcia C2}) and  (\ref{eq_01})  $N^{ref}$ is
the number of pairs without BEC and
$p_{\mu_{i}}(i = 1,\ 2)$ are four-momenta of produced particles, $M
= \sqrt {(p_1+p_2)^{2}_{\mu}}$ is the invariant mass of the pair of
bosons. For reference sample, $N^{ref}(Q)$, the like-sign pairs from
different events can be used. It is commonly assumed that the maximum
of two-particle BEC function $C^{ii}_2(Q)$ is 2 for $\vec p_{1} =
\vec p_{2}$ if no any distortion and final state interactions are
taking into account.

In general, the shape of BEC $C_2(Q)$ function is model dependent.
The most simple form of Goldhaber-like parameterization for $C_2(Q)$
[12] has been used for data fitting:
\begin{equation}
 \label{c2_aleph}
 C_2(Q)=C_0\cdot (1+\lambda e^{-Q^2R^2})\cdot (1+\varepsilon Q) ,
\end{equation}
where $C_0$ is the normalization factor, $\lambda$ is so-called the
chaoticity strength factor, meaning $\lambda =1$ for fully
chaotic and $\lambda =0$ for fully coherent sources; the
parameter $R$ is interpreted as a radius of the particle source,
often called as the "correlation radius", and assumed to be
spherical in this parameterization.
The linear term in (\ref{c2_aleph}) is often supposed to be account for long-range correlations
outside the region of BEC. However, the origin of these long-range correlations
as well as the value of $\epsilon$  are unknown yet.
Note that distribution of, e.g., pions and kaons
can be far from isotropic, usually concentrated in narrow jets, and
further complicated by the fact that the light particles with masses less
than 1 GeV often come
from decays of long-lived heavier resonances and also are under the
random chaotic interactions caused by other fields in the thermal
bath. In the parameterization (\ref{c2_aleph}) all of these problems
are embedded in the random chaoticity parameter $\lambda$.

We obtained the $C_2(Q)$ function within $QFT_\beta$ approach [3] in the form:
\begin{equation}
\label{c2_Kozlov}
 C_2(Q)=\xi(N)\cdot \Bigl[ 1+ \frac{2\alpha}{(1+\alpha)^2}\
 \sqrt{\tilde\Omega(Q)}+ \frac{1}{(1+\alpha)^2}\
 \tilde\Omega(Q) \Bigr] \cdot F(Q, \Delta x) ,
\end{equation}
where $\xi(N)$ depends on the multiplicity $N$ as
\begin{equation}
\label{eq_02}
 \xi(N)= \frac{\langle{N (N-1)}\rangle}{\langle N\rangle^2} .
\end{equation}
The consequence of the Bogolyubov's principle of weakening of
correlations at large distances [13] is given by the function $F(Q,\Delta x)$
of weakening of correlations at large spread of relative position $\Delta x$
\begin{equation}
\label{e021}
 F(Q, \Delta x) = \frac{f(Q,\Delta x)}{f(p_{1})\cdot
 f(p_{2})} = 1 + r_{f}\,Q + \ldots
\end{equation}
normalized as $F(Q, \Delta x = \infty) = 1$. Here, $f(Q,\Delta x)$ is the
two-particle distribution function with $\Delta x $, while $f(p_{i})$ are
one-particle probability functions with $i=1,2$;  $r_{f}$ is a measure
of weakening of correlations with $\Delta x $: $r_{f} \rightarrow 0$ as $\Delta
x\rightarrow \infty $.

The important parameter $\alpha$ in (\ref{c2_Kozlov}) summarizes our
knowledge of other than space-time characteristics of the particle
emitting source.

The $\tilde\Omega(Q)$ in (\ref{c2_Kozlov}) has the following
structure in momentum space
\begin{equation}
\label{e300}
 \tilde\Omega (Q)=\Omega (Q)\cdot\gamma (n) ,
\end{equation}
where
\begin{equation}
 \label{e31}
 \Omega (Q) = \exp (-\Delta_{p\Re}) =
 \exp \left [-(p_1-p_2)^{\mu}\,\Re_{\mu\nu}\, (p_1-p_2)^{\nu}\right ]
\end{equation}
is the smearing smooth dimensionless generalized function,
$\Re_{\mu\nu}$ is the (nonlocal) structure tensor of the space-time
size (BEC formation domain), and it defines the spherically-like
domain of emitted (produced) particles.

The function $\gamma (n)$ in (\ref{e300}) reflects the quantum features of BEC
pattern and is defined as
\begin{equation}
\label{eq_032}
 \gamma (n) = \frac{{n^2 (\bar \omega )}}{{n(\omega )\ n(\omega
 ')}} ,\ \
 n(\omega ) \equiv  n(\omega ,\beta ) =
 \frac{1}{{e^{(\omega  - \mu )\beta} - 1 }} ,\ \
 \bar\omega  = \frac{{\omega  + \omega '}}{2} ,
\end{equation}
where $n(\omega,\beta )$ is the mean value of quantum numbers for BE
statistics particles with the energy $\omega$ and the chemical potential $\mu$
in the thermal bath
with statistical equilibrium at the temperature $T= 1/\beta$. The
following condition $\sum_{f} n_{f}(\omega,\beta) = N$ is evident,
where the discrete index $f$ reflects the one-particle state $f$.

Note that it is commonly assumed for a long time that there are no
correlation effects among nonidentical particles (e.g., among different
charged particles). This assumption is often used in
normalizing the experimental data on $C_{2}^{ii}$ with respect to
$C_{2}^{ij}$. In the absence of interference or correlation effects
between, e.g., $\pi ^{+}$ and $\pi ^{-}$ mesons it is supposed that
$C_{2}^{+-} = 1$.

In terms of time-like $R_{0}$, longitudinal $R_{L}$ and transverse
$R_{T}$ components of the space-time size $R_{\mu}$ the distribution
$\Delta^{ij}_{p\Re}$ looks like ($i, j =+,\, -,\, 0$)
\begin{equation}
\label{e33}
 \Delta^{ij}_{p\Re}\rightarrow \Delta^{ij}_{pR} = (\Delta p^{0})^2 R^{2}_{0}
 + (\Delta p^{L})^2 R^{2}_{L} +
 (\Delta p^{T})^2 R^{2}_{T} .
\end{equation}
Seeking for simplicity one has ($R_{L}=R_{T}=R$)
\begin{equation}
\label{e344}
 \Delta^{ii}_{pR} = (p^{0}_{1}-p^{0}_{2})^{2}R^{2}_{0} +
 (\vec p_{1} - \vec p_{2})^{2} \vec R^{2}
\end{equation}
for like-sign charge bosons, while
\begin{equation}
\label{e355}
 \Delta^{ij}_{pR} = (p^{0}_{1}+p^{0}_{2})^{2}R^{2}_{0} +
 (\vec p_{1} + \vec p_{2})^{2} \vec R^{2}
\end{equation}
for different charge particles.

Obviously, the BEC effect with $\Omega^{ij}= \exp (-\Delta^{ij}_{pR})$
is smaller than that defined by $\Omega^{ii}= \exp (-\Delta^{ii}_{pR})$.
The distribution $\Omega^{ij}$ gives rise
to an off-correlation pattern between different charge particles.
The evidence of $C^{ij}_{2}$ correlation represents a
quantum-statistical correlation between a particle and an
antiparticle. Since we did not follow special assumptions on the
quantum operator level for $C_{2}$ from the initial stage, it may
correspond to a physically real and observable effect.
This pattern may lead to a new squeezing state of
correlation region. We obtain that within the $QFT_{\beta}$ the BEC
is more generally sensitive to particle-antiparticle correlations
than it would be expected from the two-particle (symmetrized) wave
function which never leads to such the correlations.

\section { Green's function}

In this paper, we would like to focus on the role of the particle mass, which
 influences the correlations between particles. To explore this problem,
one must derive the memory history of evolution of particles produced in
high energy collisions using the general properties of QFT at finite temperature.

We consider the thermal scalar complex fields  $\Phi(x)$ that correspond to
$\pi^{\pm}$  mesons with the standard definition of the Fourier transformed propagator
$F[\tilde G(p)]$
\begin{equation}
F[\tilde G(p)]= G(x-y) = Tr\left\{T[\Phi(x)\Phi(y)]\rho_{\beta}\right\},
\label{e5}
\end{equation}
with $\rho_{\beta}= e^{-\beta H}/Tr e^{-\beta H}$ being the density matrix of a local
system in equilibrium at temperature $T=\beta^{-1}$ under the Hamiltonian $H$.

We consider the interaction of $\Phi(x)$ with the external scalar field
given by the potential $U$. In contrast to an electromagnetic field, this potential
is a scalar one, but it is not a component of the four-vector. The Lagrangian
density can be written
$$ L(x) = \partial_{\mu}\Phi^{\star}(x)\partial^{\mu}\Phi(x) -
(m^2 + U)\Phi^{\star}(x)\Phi(x) $$
and the equation of motion is
\begin{equation}
(\nabla^2 + m^2)\Phi(x) = -J(x),
\label{e7}
\end{equation}
where $J(x) = U\Phi(x)$ is the source density operator. A simple model like this allows one
to investigate the origin of the
unstable state
of the thermalized equilibrium in a nonhomogeneous external field under the influence
of  source density operator $J(x)$. For example, the source can be considered
as $\delta$-like generalized function $J(x)=\tilde\mu\,\rho(x,\epsilon)\Phi(x)$ in which
$\rho(x,\epsilon)$ is a $\delta$-like succession giving the $\delta$-function as
$\epsilon\rightarrow 0$ (where $\tilde\mu$ is some massive parameter). This model is useful
because the $\delta$-like potential $U(x)$ provides the model conditions
for  restricting the particle emission domain (or the deconfinement region). We suggest the
following form:
$$ J(x) = - \Sigma(i\partial_{\mu})\,\Phi(x) + J_{R}(x), $$
where the source $J(x)$ decomposes into a regular systematic motion part
$\Sigma(i\partial_{\mu})\,\Phi(x)$ and the random source $J_{R}(x)$. Thus, the
equation of motion (\ref{e7}) becomes
$$ [\nabla^2 + m^2 - \Sigma(i\partial_{\mu})]\Phi(x) = -J_{R}(x), $$
and the propagator satisfies the following equation:
\begin{equation}
[-p^{2}_{\mu} +m^2 -\tilde\Sigma(p_{\mu})]\tilde G(p_{\mu}) =1.
\label{e10}
\end{equation}
The random noise is introduced with a random operator
$\eta(x) = - m^{-2}\,\Sigma(i\partial_{\mu})$, for that the equation of motion
looks like:
\begin{equation}
 \{\nabla^2 + m^2[1+\eta (x)]\}\Phi(x) = -J_{R}(x).
\label{e9}
\end{equation}

We assume that $\eta (x)$  varies stochastically with the certain  correlation function
(CF), e.g.,  the Gaussian CF
$$\langle\eta(x)\,\eta(y)\rangle = C\exp(-z^{2}\mu^{2}_{ch}),\,\,\, z= x-y ,$$
where $C$ is the strength of the noise described by the distribution function
$\exp(-z^{2}/L^{2}_{ch})$ with  $L_{ch}$ being the noise characteristic
scale. Both $C$ and $\mu_{ch}$ define the influence of the (Gaussian)
noise on the correlations between particles that "feel" an action of an environment.
The solution of Eq. (\ref{e9}) is
\begin{equation}
\Phi(x) = -\int dy\, G (x,y)\,J_{R}(y),
\label{e99}
\end{equation}
where the Green's function obeys the Eq.
$$\{\nabla^2 + m^2[1+\eta (x)]\}G(x,y) = \delta(x-y).$$

The final aim might having been to find the solution of Eq. (\ref{e99}), and then
average it over random operator $\eta (x)$. Note that the operator
$M(x) =\nabla^2 + m^2[1+\eta (x)]$ in the causal Green's function
$$G(x,y) = \frac{1}{M(x) +i\,o}\delta (x-y)$$ is not definitely positive.
However, we shall formulate another approach, where the random force influence
is introduced on the particle operator level.

We  introduce the general non-Fock representation
in the form of the operator generalized functions
\begin{equation}
b(x) = a(x) + r(x),
\label{e11}
\end{equation}
\begin{equation}
b^{+}(x) = a^{+}(x) + r^{+}(x),
\label{e12}
\end{equation}
where the operators $a(x)$ and $a^{+}(x)$ obey the canonical commutation
relations (CCR):
$$ [a(x),a(x^{\prime})]= [a^{+}(x),a^{+}(x^{\prime})]= 0, $$
$$ [a(x),a^{+}(x^{\prime})]= \delta (x-x^{\prime}). $$
The operator-generalized functions $r(x)$ and $r^{+}(x)$ in (\ref{e11}) and
(\ref{e12}), respectively, include random features
describing the action of the external forces.

Both $b^{+}$ and $b$ obviously define the CCR representation. For each function
$f$ from the space $S(\Re_{\infty})$ of smooth decreasing functions, one can
establish new operators $b(f)$ and $b^{+}(f)$
$$ b(f) = \int f(x) b(x) dx = a(f) +\int f(x) r(x) dx, $$
$$ b^{+}(f) = \int \bar f(x) b^{+}(x) dx = a^{+}(f) +\int\bar f(x) r^{+}(x) dx. $$
The transition from the operators $a(x)$ and $a^{+}(x)$ to $b(x)$ and $b^{+}(x)$,
obeying those commutation relations as $a(x)$ and $a^{+}(x)$, leads
to linear canonical representations.

\section {Evolution equation}




Referring to [3] for details, let us recapitulate here the
main points of our approach in the quantum case: the collision process
produces a number of
particles, out of which we select only one (we assume for simplicity that
we are dealing only with identical bosons) and describe it by stochastic
operators $b(\vec{p},t)$ and $b^{+}(\vec{p},t)$, carrying the features of
annihilation and creation operators, respectively.
The rest of the particles
are then assumed to form a kind of heat bath, which remains in
an equilibrium characterized by a temperature $T$ (one of our parameters).
We also allow for some external (relative to the above heat bath)
influence on our system.
The time evolution of such a system is then assumed to be given by a
Langevin-type equation [3] for  stochastic operator $b(\vec{p},t)$
\begin{equation}
i\partial_t b(\vec{p},t) =  A(\vec{p},t) + F(\vec{p},t) + P
\label{e20}
\end{equation}
(and a similar conjugate equation for $b^{+}(\vec{p},t)$). We assume
an asymptotic free undistorted operator  $a(\vec{p},t)$, and that the deviation
from the asymptotic free state is provided by the random operator
$r(\vec{p},t)$: $a(\vec{p},t)\rightarrow b(\vec{p},t) = a(\vec{p},t) +
r(\vec{p},t)$. This means, e.g., that the particle density number
(a physical number) ${\langle n(\vec{p},t)\rangle}_{ph} =
\langle n(\vec{p})\rangle + O (\epsilon)$, where ${\langle n(\vec{p},t)\rangle}_{ph}$
means the expectation value of a physical state, while ${\langle n(\vec{p})\rangle}$
denotes that of an asymptotic state. If we ignore the deviation
from the asymptotic state in equilibrium, we obtain an ideal fluid.
One otherwise has to consider the dissipation term;  this is why
 we use the Langevin scheme to derive the evolution equation,
 but only on the quantum level. We derive the evolution equation in an integral
form that reveals the effects of thermalization.

 Equation (\ref{e20}) is supposed to model all aspects of the hadronization processes
(or even deconfinement). The combination $A(\vec{p},t)+ F(\vec{p},t)$ in the r.h.s of
(\ref{e20}) represents the
so-called {\it Langevin force} and is therefore responsible for the
internal dynamics of particle emission, as the memory term $A$ causes
dissipation and is related to stochastic dissipative forces [3]
$$ A(\vec{p},t) = \int^{+\infty}_{-\infty}\! d\tau K(\vec{p},t-\tau)
b(\vec{p},\tau) $$
with  $K(\vec{p},t)$ being the kernel operator describing the
virtual transitions from one (particle) mode to another.
At any dependence of the field operator $K$ on the time, the function
$A(\vec{p},t)$ is defined by the behavior of the system at the precedent moments.
The operator $F(\vec{p},t)$  in (\ref{e20}) is responsible for the action of
a heat bath of absolute temperature $T$ on a particle in the heat bath, and
under the appropriate circumstances is given by
$$ F(\vec{p},t) =
\int^{+\infty}_{-\infty}\!\frac{d\omega}{2\pi}\psi(p_{\mu})\hat{c}(p_{\mu})
e^{-i\omega t} . $$
The heat bath is represented by an ensemble of coupled oscillators,
each described by the operator $\hat{c}(p_{\mu})$ such
that $\left[\hat{c}(p_{\mu}),\hat{c}^{+}(p'_{\mu})\right] =
\delta^4(p_{\mu}-p'_{\mu})$, and is characterized by the noise spectral function
$\psi(p_{\mu})$ [3]. Here, the only statistical assumption is that the heat bath
is canonically distributed. The oscillators are coupled to a particle, which is
in turn acted upon by an outside force.
Finally, the constant term $P$ in (\ref{e20}) (representing {\it an external source}
term in the Langevin equation) denotes a possible influence of
some external force. This force
would result, e.g., in a strong ordering of phases leading
therefore to a coherence effect.

The solution of equation (\ref{e20}) is given in $S(\Re_{4})$ by
\begin{equation}
\tilde b(p_{\mu}) = \frac{1}{\omega - \tilde K(p_{\mu})}\,
[\tilde F(p_{\mu}) + \rho (\omega_{P},\epsilon)],
\label{e23}
\end{equation}
where $\omega$ in $\rho (\omega,\epsilon)$ was replaced by  new scale
$\omega_{P} = \omega/P$.
It should be stressed that the term containing
$\rho (\omega_{P},\epsilon)$ as $ \epsilon\rightarrow 0$ yields the general solution to
Eq. (\ref{e20}). Notice that the distribution $\rho (\omega_{P},\epsilon)$  indicates
the continuous character of the spectrum, while the arbitrary small quantity
$\epsilon$ can be defined by the special physical conditions or the physical
spectra. On the other hand, this $\rho (\omega_{P},\epsilon)$ can be understood as
temperature-dependent succession
$\rho (\omega,\epsilon)=\int dx\,exp (i\omega - \epsilon)x \rightarrow \delta (\omega)$,
in which  $\epsilon\rightarrow \beta^{-1}$.
Such a succession yields the restriction on the $\beta$-dependent
second term in the solution (\ref{e23}), where at small enough $T$ there is
a narrow peak at $\omega = 0$.

 From the scattering matrix point of view, the solution (\ref{e23}) has the following
physical meaning: at a sufficiently  outgoing past and future, the fields described by the
operators $\tilde a(p_{\mu})$ are free and  the initial and the final states
of the dynamic system are thus characterized by  constant amplitudes.
Both  states, $\varphi (-\infty)$ and $\varphi (+\infty)$, are related to
one another by an operator $S(\tilde r)$ that transforms  state
$\varphi (-\infty)$ to  state $\varphi (+\infty)$ while depending on the behaviour
of $\tilde r(p_{\mu})$:
$$\varphi (+\infty)  = S(\tilde r)\varphi (-\infty).$$
In accordance with this definition, it is natural to identify $S(\tilde r)$ as the
scattering matrix in the case of arbitrary sources that give rise to the intensity of
$\tilde r$.

Based on QFT point of view, relation (\ref{e11}) indicates the
appearance of the terms containing nonquantum fields that are characterized by
the operators $\tilde r(p_{\mu})$.
Hence, there are terms with $\tilde r$ in the matrix elements, and
these $\tilde r$ cannot be realized via real particles. The operator function
$\tilde r(p_{\mu})$ could be considered as the limit on an average value of some quantum
operator (or even a set of operators) with an intensity that increases to infinity.
The later statement can be visualized in the following mathematical representation:
$$\tilde r(p_{\mu}) = \sqrt {\alpha\,\Xi (p_{\mu},p_{\mu})},\,\,
\Xi (p_{\mu},p_{\mu}) = {\langle\tilde a^{+}(p_{\mu})\, \tilde a(p_{\mu})\rangle}_{\beta} ,
 $$
where $\alpha$ is the coherence (chaotic) function that gives the strength of the average
$\Xi (p_{\mu},p_{\mu})$.

In principal, interaction with the fields described by $\tilde r$ is provided by
the virtual particles, the propagation process of which is given by the potentials
defined by the $\tilde r$ operator function.

The condition $M_{ch}\rightarrow 0$ (or $\Omega_{0}(R)\sim\frac{1}{M^{4}_{ch}}\rightarrow\infty$)
in the representation
$$\lim_{p_{\mu}\rightarrow p^{\prime}_{\mu}}\Xi (p_{\mu},p^{\prime}_{\mu}) =
\lim_{Q^{2}\rightarrow 0} \Omega_{0}(R)\,n(\bar\omega,\beta)\exp (-q^{2}/2)\rightarrow
\frac{1}{M^{4}_{ch}}\,n(\omega,\beta), $$
with
$$ \Omega_{0}(R)=\frac{1}{\pi^2}\,R_{0}\,R_{L}\,R^{2}_{T}
 $$
means that the role of the arbitrary source characterized by the operator function
$\tilde r(p_{\mu})$ in $\tilde b(p_{\mu})= \tilde a(p_{\mu}) + \tilde r(p_{\mu})$ disappears.

\section{ Green's function and kernel operator}

Let us go to the thermal field operator $\Phi(x)$ by means of the linear combination
of the frequency parts $\phi^{+}(x)$ and $\phi^{-}(x)$
\begin{equation}
\Phi (x) = \frac{1}{\sqrt{2}}\,\left [\phi^{+}(x) +\phi^{-}(x)\right ]
\label{e24}
\end{equation}
composed of the operators $\tilde b(p_{\mu})$ and $\tilde b^{+}(p_{\mu})$ as the
solutions of equation (\ref{e20}) and conjugate to it, respectively:
$$ \phi^{-} (x) = \int \frac{d^{3}\vec{p}}{(2\pi)^{3} 2 (\vec p^{2} +m^{2})^{1/2}}
\tilde b^{+}(p_{\mu})\,e^{ipx}, $$
$$\phi^{+} (x) = \int \frac{d^{3}\vec{p}}{(2\pi)^{3} 2 (\vec p^{2} +m^{2})^{1/2}}
\tilde b(p_{\mu})\,e^{-ipx}. $$
The function $\Phi (x)$ obeys the commutation relation
$$[\Phi (x),\Phi (y)]_{-} = - i D(x) $$
with [14]
$$ D(x)= \frac{1}{2\,\pi}\,\epsilon(x^{0})\,\left [\delta(x^2) -
\frac{m}{2\,\sqrt{x^{2}_{\mu}}}\,\Theta(x^2)\,J_{1}\left (m\sqrt{x^2_{\mu}}\right )\right ], $$
where $\epsilon(x^{0})$ and $\Theta(x^2)$ are the standard unit and the step functions,
respectively, while $ J_{1}(x)$ is the Bessel function. On the mass-shell,
$D(x)$ becomes [14]
$$ D(x)\simeq \frac{1}{2\,\pi}\,\epsilon(x^{0})\,\left [\delta(x^2) -
\frac{m^2}{4}\,\Theta(x^2) \right ]. $$

One can easily find two equations of motion for the Fourier transformed operators
$\tilde b(p_{\mu})$ and $\tilde b^{+}(p_{\mu})$ in $S(\Re_{4})$
\begin{equation}
[\omega - \tilde K(p_{\mu})]\tilde b(p_{\mu}) = \tilde F(p_{\mu}) + \rho(\omega_{P},\epsilon),
\label{e27}
\end{equation}
\begin{equation}
[\omega - \tilde K^{+}(p_{\mu})]\tilde b^{+}(p_{\mu}) =
\tilde F^{+}(p_{\mu}) + \rho^{\star}(\omega_{P},\epsilon),
\label{e28}
\end{equation}
which are transformed into new equations for the frequency parts
$\phi^{+} (x)$ and $\phi^{-} (x)$ of the field operator $\Phi (x)$ (\ref{e24})
\begin{equation}
i\partial_{0}\phi^{+} (x) + \int_{\Re_{4}} K(x-y)\,\phi^{+} (y)dy = f(x)
\label{e29}
\end{equation}
\begin{equation}
- i\partial_{0}\phi^{-} (x) + \int_{\Re_{4}} K^{+}(x-y)\,\phi^{-} (y)dy = f^{+}(x),
\label{e30}
\end{equation}
where
$$f(x) =\int\frac{d^{3}\vec{p}}{(2\pi)^{3}\,(\vec p^{2} +m^{2})^{1/2}} [\tilde F(p) +\rho(\omega_{P},\epsilon)]
e^{-ipx},$$
and
$$f^{+}(x) =\int\frac{d^{3}\vec{p}}{(2\pi)^{3}\,(\vec p^{2} +m^{2})^{1/2}} [\tilde F^{+}(p)
+\rho^{\star}(\omega_{P},\epsilon)] e^{ipx}.$$

Here, the field components $\phi^{+}(x)$ and  $\phi^{-}(x)$ are under the effect
of the  nonlocal formfactors $K(x-y)$ and $K^{+}(x-y)$, respectively. In general,
these formfactors can
admit the description of locality for nonlocal interactions.

At this stage, it must be stressed that we have new generalized evolution
Eqs. (\ref{e29}) and (\ref{e30}), which retain  the general
features of the propagating and
interacting of the quantum fields with mass $m$ that are in the heat bath (reservoir)
and are chaotically distorted by  other fields. For  further analysis,
let us rewrite the  Eqs. (\ref{e29}) and (\ref{e30}) in the following form:
\begin{equation}
i\partial_{0}\phi^{+} (x) +  K(x)\star\phi^{+} (x) = f(x),
\label{e34}
\end{equation}
\begin{equation}
- i\partial_{0}\phi^{-} (x) +  K^{+}(x)\star\phi^{-} (x) = f^{+}(x),
\label{e35}
\end{equation}
where $A(x)\star B(x)$ is the convoluted function of the generalized functions
$A(x)$ and $B(x)$.
Applying the direct Fourier transformation to both sides of Eqs.
(\ref{e34}) and (\ref{e35}) with the following properties of the
Fourier transformation
$$ F[K(x)\star\phi^{+} (x) ] = F[K(x)]F[\phi^{+} (x)],$$
we  get two equations
\begin{equation}
[p^{0} + \tilde K(p_{\mu})]\tilde\phi^{+}(p_{\mu}) = F[f(x)],
\label{e388}
\end{equation}
\begin{equation}
[- p^{0} + \tilde K^{+}(p_{\mu})]\tilde\phi^{-}(p_{\mu}) = F[f^{+}(x)].
\label{e399}
\end{equation}
Multiplying Eqs. (\ref{e388}) and (\ref{e399}) by $- p^{0} + \tilde K^{+}(p_{\mu})$
and $ p^{0} + \tilde K(p_{\mu})$, respectively, we find
\begin{equation}
[- p^{0} + \tilde K^{+}(p_{\mu})][p^{0} + \tilde K(p_{\mu})]\tilde\Phi(p_{\mu}) = T(p_{\mu}),
\label{e400}
\end{equation}
where
$$ T(p_{\mu}) = [- p^{0} + \tilde K^{+}(p_{\mu})]F[f(x)]+
 [p^{0} + \tilde K(p_{\mu})]F[f^{+}(x)]. $$
We are now  at the stage of the main strategy:  we have to identify the field
$\Phi (x)$ introduced in Eq. (\ref{e5}) and the field $\Phi (x)$ (\ref{e24}) built up of the
fields $\phi^{+}$ and $\phi^{-}$ as the solutions of  generalized
Eqs. (\ref{e29}) and (\ref{e30}). The next step is our requirement that
Green's function
$\tilde G(p_{\mu})$ in Eq. (\ref{e10}) and  the function $ \Gamma(p_{\mu})$,
that satisfies Eq. (\ref{e400})
\begin{equation}
[- p^{0} + \tilde K^{+}(p_{\mu})][p^{0} + \tilde K(p_{\mu})]\tilde\Gamma(p_{\mu}) = 1,
\label{e42}
\end{equation}
must be equal to each other, where the full Green's function $\tilde G(p^2, g^2, m^2)$
\begin{equation}
 \tilde G(p_{\mu})\rightarrow \tilde G(p^2, g^2, m^2)\simeq \frac{1 - g^2\,\xi(p^2, m^2)}
{m^2 - p^2 -i\epsilon}
\label{e43}
\end{equation}
has the same pole structure at $p^{2} = m^{2}$ as the free Green's function [14]
with $g$ being the scalar coupling constant and $\xi$ is the one-loop correction
of the scalar field. The dimensioneless function $1 - g^2\,\xi(p^2, m^2)$ is finite
at $p^{2} = m^{2}$.

We define the operator kernel $\tilde K(p_{\mu})$ in (\ref{e27}) from
the condition of the nonlocal coincidence of the Green's function $\tilde G(p_{\mu})$
in Eq. (\ref{e10}), and the thermodynamic function $\tilde\Gamma(p_{\mu})$
from (\ref{e42}) in $S(\Re_{4})$
$$ F[\tilde G(p_{\mu}) - \tilde\Gamma(p_{\mu})] =0.$$

We can easily derive the kernel operator
$\tilde K(p_{\mu})$ in the form
\begin{equation}
\tilde K^{2}(p) = \frac{m^2 + \vec p^2 -g^2\xi(p^2, m^2)\,p^{0^{2}}}{1- g^2\xi(p^2, m^2)}
\label{e44}
\end{equation}
where [14]
$$ \xi(m^2) = \frac{1}{96\,\pi^2\,m^2}\left (\frac{2\,\pi}{\sqrt{3}} -1\right ),
\,\,\, p^2\simeq m^2, $$
and
$$ \xi(p^2, m^2) = \frac{1}{96\,\pi\,m^2}\left (i\, \sqrt {1-\frac{4\,m^2}{p^2}}
+ \frac{\pi}{\sqrt{3}}\right ),
\,\,\, p^2\simeq 4 m^2. $$
The ultraviolet behaviour at $ \vert p^2\vert >> m^2$ leads to

$$ \xi(p^2, m^2) \simeq  \frac{-1}{32\,\pi^{2}\,p^2}\left [\ln\frac{\vert p^2\vert}{m^2} -
\frac{\pi}{\sqrt{3}} - i\,\pi\Theta (p^2)\right ]. $$

\section{Stochastic forces scale}

In paper  [4] it has been emphasized that two different scale
parameters are in the model which we consider
here. One of them is the so-called "correlation radius" $R$ introduced
in (\ref{c2_aleph}) and (\ref{c2_Kozlov}) with (\ref{e300}) and
(\ref{e33}), (\ref{e344}), (\ref{e355}). In fact,
this $R$-parameter gives the pure size of the particle emission
source without the external distortion and interaction coming from
other fields. The other (scale) parameter is the stochastic
scale $L_{st}$ which carries  the dependence of the particle mass,
the $\alpha$-coherence degree and what is very important --- the
temperature $T$-dependence:
\begin{equation}
\label{e36}
 L_{st}={\left[\frac{1}{\alpha(N)\,{\vert p^{0}-\tilde K(p)\vert}^{2}\,
 \bar n(m,\beta)}\right ]}^{\frac{1}{5}}.
\end{equation}
It turns out that this scale $L_{st}$ defines the range of stochastic forces acting the particles
in the emission source.
This effect is given by $\alpha
(N)$-coherence degree which can be estimated from the experiment within the two-particle BE
correlation function $C_{2}(Q)$ as $Q$ close to zero, $C_{2}(0)$,
at fixed value of mean multiplicity $\langle N\rangle$:
\begin{equation}
 \label{e37}
\alpha (N)\simeq \frac{2-\bar C_{2}(0) + \sqrt {2-\bar
C_{2}(0)}}{\bar C_{2}(0)-1},\,\,\, \bar C_{2}(0)= C_{2}(0)/\xi(N) .
\end{equation}
In formula (\ref{e36}) $\bar n(m,\beta)$ is the thermal
relativistic particle number density
\begin{equation}
 \label{e38}
\bar n(m,\beta) = 3\int\frac{d^{3}\vec
p}{(2\,\pi)^{3}}\,n(\omega,\beta)=3\frac{\mu^2 +m^2}
{2\,\pi^2}\,T\,\sum_{l=1}^{\infty}\frac{1}{l}K_{2}\left
(\frac{l}{T}\sqrt{\mu^2 + m^2}\right) ,
\end{equation}
where $K_{2}$ is the modified Bessel function. For definite calculations we consider
correlations between charge pions. The result can be extended to heavy particles case,
e.g., for charge and neutral gauge bosons that is essential program for the LHC.
The stochastic scale $L_{st}$ tends to infinity in case of particles are on mass-shell,
i.e., $\vert p^{0}-\tilde K(p)\vert\rightarrow 0$ which enters the $L_{st}'s$
denominator (\ref{e36}). However, $L_{st}$ will be bounded due to stochastic forces
acting the particles where
$${\vert p^{0}-\tilde K(p)\vert}^{2}\simeq \Delta\epsilon^{2}_{p}=
\epsilon^{2}_{p}\left \vert \frac{p^{0}}{\epsilon_{p}} - 1 - \frac{g^{2}\xi(p^{2}, m^{2})}{2}
\left ( 1-\frac{p^{0^{2}}}{\epsilon^{2}_{p}}\right )\right\vert ^{2},\,\,
\epsilon_{p} =\sqrt {m^{2}+\vec p^{2}}$$
as $g^{2}\xi(p^{2}, m^{2}) <1$.

Within our aim to
explore the correlation between charged pions, $L_{st}$ has the form
\begin{equation}
 \label{e39}
L_{st}\simeq {\left [\frac{e^{\sqrt {\mu^{2}+m^{2}}/T}}
{3\, \alpha (N)\,\Delta\epsilon^{2}_{p}\,{(\mu^{2}+m^{2})}^{3/4}\,
{\left (\frac{T}{2\,\pi}\right)}^{3/2}\,
\left (1+\frac{15}{8}\frac{T}{\sqrt {\mu^{2}+m^{2}}}\right )}\right ]}^{\frac{1}{5}} ,
\end{equation}
where the condition $l\,\beta\sqrt{m^{2}+\mu^{2}} >1$  for any integer $l$
 in (\ref{e38}) was taken into account. The only lower temperatures will drive
$L_{st}$ within the formula (\ref{e39}) even if $\mu =0$ and $l=1$ with the
condition $T < m$.
Note that the
condition $\mu < m$ is a general restriction in the relativistic
"Bose-like gas", and $\mu = m$ corresponds to the Bose-Einstein
condensation.

For large enough $T$ no the dependence of the chemical potential $\mu$ is found
for $L_{st}$:
\begin{equation}
 \label{e40}
L_{st}\simeq
{\left [\frac{\pi^{2}}{3\,\zeta (3)\, \alpha(N)\,\Delta\epsilon^{2}_{p}\,T^{3}}\right ]}^{\frac{1}{5}} ,
\end{equation}
where the condition $T > l\sqrt {\mu^{2} + m^{2}}, l=1,2,...$ is taken into account.
The origin of formula (\ref{e40}) comes from
\begin{equation}
 \label{e41}
 \bar n (m,\beta)\rightarrow \bar n(\beta) = \frac{3\,T^{3}}{\pi^{2}}\,\zeta (3)
\end{equation}
where neither a pion mass $m$- nor $\mu$- dependence occurred;
$\zeta (3)= \sum^{\infty}_{l=1} l^{-3} = 1.202$ is the zeta-function with the argument $3$.
For high momentum pions ($p^2\simeq 4 m^2$) the actual mass-dependence occurred for $L_{st}$:
\begin{equation}
 \label{e399}
L_{st}\simeq {\left [\frac{e^{\sqrt {\mu^{2}+m^{2}}/T}}
{3\, \alpha (N)\,m^{2}\,{(\mu^{2}+m^{2})}^{3/4}\,
{\left (\frac{T}{2\,\pi}\right)}^{3/2}\,
\left (1+\frac{15}{8}\frac{T}{\sqrt {\mu^{2}+m^{2}}}\right )}\right ]}^{\frac{1}{5}} ,
\end{equation}
at low T, and
\begin{equation}
 \label{e400}
L_{st}\simeq
{\left [\frac{\pi^{2}}{3\, \zeta (3)\,\alpha(N)\,m^{2}\,T^{3}}\right ]}^{\frac{1}{5}} ,
\end{equation}
at high temperatures, if $g^{2}\xi(m^{2}) <<1$, $\xi(m^{2})\sim O(0.01/m^{2})$ and
$(\vec p^{2}/4m^{2}) <<1$ are valid in both temperature regime cases.
Formula (\ref{e41}) reproduces the $\sim T^{3}$ behavior which is the same as the
thermal distribution (in terms of density) for a gas of free relativistic massless
particles. Such a behavior is expected anyway in high temperature limit if the particles
can be considered as asymptotically free in that regime.

Actually, the increasing of $T$ leads to squeezing of $L_{st}$, and
$L_{st}(T=T_{0})= R$ at some effective temperature  $T_{0}$.
The higher temperatures, $T > T_{0}$,
satisfy to more squeezing effect and at the critical temperature
$T_{c}$ the scale $L_{st}(T=T_{c})$ takes its minimal value.
Obviously $T_{c}$ defines the phase transition where the
deconfinement will occur. Since all the masses tend to zero (chiral
symmetry restoration) and $\alpha\rightarrow 0$ at $T>T_{c}$ one
should expect the sharp expansion of the region with
$L_{st}(T>T_{c})\rightarrow \infty$. The following condition $\tilde
n(m,\beta)\cdot v_{\pi} =1$ provides the phase transition
(transition from hadronizing phase to deconfinement one) with the
volume $v_{\pi} = (4\,\pi\,r^{3}_{\pi}/3)$, where $r_{\pi}$ is the
pion charge radius. Actually, the temperature of phase transition essentially
depends on the charge (vector) radius of the pion which is a fundamental
quantity in hadron physics. A recent review on $r_{\pi}$ values is presented in [15].

What we know about the source size estimation from experiments? DELPHI and L3
collaborations at LEP established that the correlation radius $R$ decreases
with transverse pion mass $m_{t}$ as $R\simeq a + b/\sqrt{m_{t}}$ for all directions
in the Longitudinal Center of Mass System (LCMS). ZEUS collaboration at HERA
did not observe the essential difference between the values of $R$ - parameter in
$\pi^{\pm}\pi^{\pm}, \, K^{0}_{s}K^{0}_{s}$ and $K^{\pm}K^{\pm}$ pairs, namely
$R_{\pi\pi} = 0.666 \pm 0.009 (stat) + 0.022 - 0.033 (syst.) fm$,
$R_{K_{s}K_{s}} = 0.61 \pm 0.08 (stat) + 0.07 - 0.08 (syst.) fm$ and
$R_{K^{\pm}K^{\pm}} = 0.57 \pm 0.09 (stat) + 0.15 - 0.06 (syst.) fm$, respectively.
The ZEUS data are in good agreement with the LEP for radius $R$.
However, no evidence for $\sqrt s$ dependence of $R$ is found.
It is evidently that more experimental data are appreciated. However, the comparison
between experiments is difficult mainly due to reference samples used and the Monte Carlo
corrections.

Finally, our theoretical results first predict the $L_{st}$ in (\ref{e39}) and
(\ref{e40}), and both mass- and temperature - dependence are obtained clearly.
 This can serve as a good approximation to  explain the LEP, Tevatron and ZEUS (HERA)
 experimental data.
 We need that the pion energies at the colliders are sufficient to carry these studies out
 (since the $\Delta\epsilon_{p}$ dependence). Careful simulation of their (pions)
 signal and background are needed. The more precisely measured pion momentum may be of some help.
 Also, determination of the final state interactions may clarify what is happening.

\section{Conclusions}
%

To summarize: we find the time dependence of the correlation function $C_2(Q)$
calculated in time-dependent external field provided by the operator
$r(\vec p, t)$ and the chaotic coherence function $\alpha(m,\beta)$.
Based on this approach we emphasize
the explanation of the dynamic origin of the coherence in
BEC, the origin of the specific shape of the correlation $C_2(Q)$
functions, and finding the dependence on the
particle energy (and the mass) due to  coherence function $\alpha$,
as seen from the $QFT_{\beta}$. Actually, the stochastic scale $L_{st}$ decreases
with the particle energy (the mass $m$).
It is already confirmed by the data of LEP, Tevatron and HERA (ZEUS) with respect to
the size of particle source.

In the framework of ${QFT}_{\beta }$ the numerical analysis of experimental data
can be carried out with a result where important parameters of ${C_2}^{--}(Q)$
and ${C_2}^{++}(Q)$ functions are retrieved (e.g., $C_{0}, R, \lambda, \epsilon, N, \alpha,
L_{st}, T$).

The correlations of non-identical particles pairs can be observed and the
corresponding $C_2$ parameters is retrieved.
The off-correlation effect is given by the space-time distribution (\ref{e355}) containing
the sum $\vec p_{1} + \vec p_{2}$, and this effect is sufficient if the factor containing
the sum $p^{0}_{1} + p^{0}_{2}$ in (\ref{e31}) is not too small. The off-correlation
effect is possible if the particle energies $p^{0}_{i} (i=1,2)$ are small enough.


Besides the fact that like, e.g.,  $\pi^{\pm}\pi^{\pm}$ BEC the correlations
$\pi^{\pm}\pi^{\mp}$ can serve as tools in the determination of parameters of
the particle source.
And besides the fact that these correlations
play a particularly important role in the detection of random
chaotic correction to BEC.

The stochastic scale $L_{st}$ decreases with increasing temperatures slowly at low
temperatures, and it decreases rather abruptly when the critical temperature is approached.

We claim that the experimental measuring of $R$ (in $fm$) can provide the precise estimation
of the effective temperature $T_{0}$ which is the main thermal character
in the particle's pair emitter source (given by the effective dimension $R$)
with the particle mass and its energy at
given $\alpha$ fixed by $C_{2}(Q=0)$ and $\langle N\rangle$. Actually, $T_{0}$ is
the true temperature in the region of multiparticle production with dimension
$R = L_{st}$, because at this temperature it is exactly the creation of two
particles occurred, and these particles obey the criterion of BEC.

We have found the  squeezing of the particle  source due to decreasing of the
correlation radius $R$ in the case of  opposite charge particles.
The off-correlated system of non-identical particles is less sensitive to the random
force influence ($\alpha$- dependence).

The results obtained in this paper can be compared with the static correlation function
(see, e.g., [16] and the references therein relevant to heavy ion collisions).

Finally, we should stress a new features of particle-antiparticle
BEC which can emerge from the data. It is a highly rewarding task to
experimental measurement of non-identical particles.

There is much to be done for $C_{2}(Q)$ investigation at hadron colliders.
The time is ripe for dedicated searches for new effects in $C_{2}(Q)$
function at hadron colliders to discover, or rule out, in particular, the $\alpha(N)$
dependence.

In conclusion, the correlations of two bosons in 4-momentum space presented
in this paper offer useful
and instructive complimentary viewpoints to theoretical and experimental works in
multiparticle femtoscopy and interferometry measurements at hadron colliders.




\end{document}